\documentclass[a4,12pt]{article}
\usepackage{float}
\usepackage{amsmath}
\def\ls{\lower0.5ex\hbox{$\buildrel >\over{\scriptstyle\sim}$}}
\def\rs{\lower0.5ex\hbox{$\buildrel <\over{\scriptstyle\sim}$}} 
\allowdisplaybreaks
\begin{document}
\pagestyle{empty} \setlength{\footskip}{2.0cm}
\setlength{\oddsidemargin}{0.5cm}
\setlength{\evensidemargin}{0.5cm}
\renewcommand{\thepage}{-- \arabic{page} --}
\def\mib#1{\mbox{\boldmath $#1$}}
\def\bra#1{\langle #1 |}  \def\ket#1{|#1\rangle}
\def\vev#1{\langle #1\rangle} \def\dps{\displaystyle}
\newcommand{\fcal}{{\cal F}}
\newcommand{\gcal}{{\cal G}}
\newcommand{\ocal}{{\cal O}}
\newcommand{\El}{E_\ell}
\renewcommand{\thefootnote}{$\sharp$\arabic{footnote}}
\newcommand{\W}{{\scriptstyle W}}
 \newcommand{\I}{{\scriptscriptstyle I}}
 \newcommand{\J}{{\scriptscriptstyle J}}
 \newcommand{\K}{{\scriptscriptstyle K}}
%
 \def\thebibliography#1{\centerline{REFERENCES}
 \list{[\arabic{enumi}]}{\settowidth\labelwidth{[#1]}\leftmargin
 \labelwidth\advance\leftmargin\labelsep\usecounter{enumi}}
 \def\newblock{\hskip .11em plus .33em minus -.07em}\sloppy
 \clubpenalty4000\widowpenalty4000\sfcode`\.=1000\relax}\let
 \endthebibliography=\endlist
 \def\sec#1{\addtocounter{section}{1}\section*{\hspace*{-0.72cm}
 \normalsize\bf\arabic{section}.$\;$#1}\vspace*{-0.3cm}}
\def\secnon#1{\section*{\hspace*{-0.72cm}
 \normalsize\bf$\;$#1}\vspace*{-0.3cm}}
 \def\subsec#1{\addtocounter{subsection}{1}\subsection*{\hspace*{-0.4cm}
 \normalsize\bf\arabic{section}.\arabic{subsection}.$\;$#1}\vspace*{-0.3cm}}
\vspace*{-1.7cm}
\begin{flushright}
$\vcenter{
\hbox{{\footnotesize OUS and TOKUSHIMA Report}}
{ \hbox{(arXiv:1511.03437)}  }
}$
\end{flushright}

\vskip 1.6cm
\begin{center}
  {\large\bf Full analysis of general non-standard $\mib{tbW}$ couplings}
\vskip 0.18cm

\end{center}

\vspace{0.9cm}
\begin{center}
\renewcommand{\thefootnote}{\alph{footnote})}
Zenr\=o HIOKI$^{\:1),\:}$\footnote{E-mail address:
\tt hioki@tokushima-u.ac.jp}\ and\
Kazumasa OHKUMA$^{\:2),\:}$\footnote{E-mail address:
\tt ohkuma@ice.ous.ac.jp}
\end{center}

\vspace*{0.4cm}
\centerline{\sl $1)$ Institute of Theoretical Physics,\ University of Tokushima}

\centerline{\sl Tokushima 770-8502, Japan}

\vskip 0.2cm
\centerline{\sl $2)$ Department of Information and Computer Engineering,}

\centerline{\sl Okayama University of Science}

\centerline{\sl Okayama 700-0005, Japan}

\vspace*{2.6cm}
\centerline{ABSTRACT}

\vspace*{0.2cm}
\baselineskip=21pt plus 0.1pt minus 0.1pt
Possible non-standard couplings which could contribute to the $t\to b W$ process are studied 
based on the effective-Lagrangian approach. The corresponding effective Lagrangian consists
of four kinds of dimension-6 effective operators, each of which has an independent coupling
constant. In this analysis, all those couplings are treated as complex numbers and constraints
on them are estimated by using recent experimental data from the LHC. We point out that
the resultant constraints on those couplings are still not that strong because contributions
from some couplings can work oppositely with each other. 
\vskip 1.5cm

\vfill

\setcounter{page}{0}
\newpage
\renewcommand{\thefootnote}{$\sharp$\arabic{footnote}}
\pagestyle{plain} \setcounter{footnote}{0}

The top quark, the mass of which is about 173 GeV, is still the heaviest particle we can
observe up to now although a new scalar indicating the Higgs boson, the last
piece of the standard model, has been discovered~\cite{Aad:2012tfa, Chatrchyan:2012xdj}.
Studying this quark from various angles will be, therefore, a quite promising approach to
new physics beyond the standard model~\cite{Atwood:2000tu,Kamenik:2011wt,Tait:2015usa}.
In particular, precise analyses of the top-quark couplings could play a crucial role
to reveal new-physics effects that might exist behind phenomena observed in collider
experiments. We will soon have more information for those studies, considering that
the Large Hadron Collider (LHC) has now re-started measuring the top-quark properties
more precisely with $\sqrt{s}=13$ TeV and a plan of luminosity upgrade~\cite{LHC}. 

In precision measurements, a sign of new-physics will appear in various observables as
deviations from the standard-model predictions, unless new (non-standard) particles are
directly discovered. Since those deviations in general arise through quantum loop effects
of non-standard particles, the effective-Lagrangian procedure
\cite{Buchmuller:1985jz}--\cite{Grzadkowski:2010es} is known as a useful way to describe
such effects. This approach enables a model-independent analysis if we construct the
effective Lagrangian using only the standard-model fields below the new-physics scale
(${{\mit\Lambda}}$). The top-quark-decay process we focus on, $t \to bW$, is
suitable for those studies because a top quark decays quickly within the perturbation
region owing to its heavy mass~\cite{Bigi:1980az,Bigi:1986jk}. 

Although many authors have already studied top-decay processes in the effective-Lagrangian
framework in order to probe possible new interactions~\cite{Kane:1991bg}--\cite{Aguilar:2015vsa},
the non-standard couplings included there have been treated as real numbers, or as partially complex
numbers, and/or only some couplings have been treated as free parameters at once fixing the others.
In addition, it has not been unusual to adopt the linear approximation in those parameters,
i.e., to neglect their quadratic (and higher-power) terms. 
Those limited analyses could be reasonable if the authors are implicitly considering some
specific models. We cannot say however that they are the most satisfactory as purely model-independent
studies. Therefore, in this short article, assuming all those non-standard couplings are complex
numbers and contribute to the top-decay process at the same time, we estimate current constraints
on them from recent experimental data without taking the linear approximation.

In our analysis, we assume that there exist no new particles at any energy less than ${{\mit\Lambda}}$. 
Based on this assumption and adopting 
the notations of our previous work~\cite{HIOKI:2011xx,Hioki:2012vn,Hioki:2014eca},
the effective Lagrangian for $t\to bW$ is expressed as
\begin{alignat}{1}\label{eq:efflag_decay}
  &{\cal L}_{tbW}  = -\frac{1}{\sqrt{2}}g 
  \Bigl[\,\bar{\psi}_b(x)\gamma^\mu(f_1^L P_L + f_1^R P_R)\psi_t(x)W^-_\mu(x) \Bigr.
  \nonumber\\
 &\phantom{========}
  +\bar{\psi}_b(x)\frac{\sigma^{\mu\nu}}{M_W}(f_2^L P_L + f_2^R P_R)
   \psi_t(x)\partial_\mu W^-_\nu(x) \,\Bigr],
\end{alignat}
where $g$ is the $SU(2)$ coupling constant, 
$P_{L/R}\equiv(1\mp\gamma_5)/2$, and the coupling parameters $f_{1,2}^{L,R}$
are defined as
\begin{alignat}{2}\label{eq:fdef}
  f_1^L&\equiv V_{tb}+C^{(3,33)*}_{\phi q}\frac{v^2}{{{\mit\Lambda}}^2},  
 & \quad  f_1^R&\equiv C^{33*}_{\phi \phi}\frac{v^2}{2{{\mit\Lambda}}^2},  \\
  f_2^L&\equiv -\sqrt{2} C^{33*}_{dW}\frac{v^2}{{{\mit\Lambda}}^2},
 & \quad f_2^R&\equiv -\sqrt{2} C^{33}_{uW}\frac{v^2}{{{\mit\Lambda}}^2} \nonumber
\end{alignat}
with $v$ being the Higgs vacuum expectation value, $V_{tb}$ being the ($tb$) element
of Kobayashi--Maskawa matrix, and $C^{(3,33),33}_{\phi q,\phi\phi,dW,uW}$ being the parameters
representing the contributions of the corresponding dimension-6 operators (see
\cite{AguilarSaavedra:2008zc}). Among those parameters, we divide $f_1^L$ into
the SM term and the rest (i.e., the non-SM term) as 
\begin{equation}
f_1^L \equiv f_1^{\rm SM}+\delta\! f_1^L,
\label{eq:f1redef}
\end{equation}
where $f_1^{\rm SM}\equiv V_{tb}$ and
$\delta\! f_1^L\equiv C^{(3,33)*}_{\phi q}{v^2}/{{{\mit\Lambda}}^2}$.
We then assume $f_1^{\rm SM} (= V_{tb})=1$ and treat $\delta\! f_1^L$, $f_1^R$, and
$f_2^{L/R}$ as complex numbers hereafter. As for the masses of the involved particles,
we take as $m_t=172.5~{\rm GeV}$, $m_b=4.8~{\rm GeV}$ and $M_W=80.4~{\rm  GeV}$.

Now, we here focus on $t\to b W$ as mentioned and assume that it is the unique top-decay
channel. The $W$-boson is produced there with one of the following helicities:
$h=0$ (longitudinal), $h=-1$ (left-handed), and $h=+1$ (right-handed), which means
there are three kinds of helicity fraction corresponding to each helicity state.
The analytical formulas of those partial decay widths are calculated by using
Eq.(\ref{eq:efflag_decay}) straightforwardly and we have confirmed that our formulas
are the same as those presented in Ref.\cite{AguilarSaavedra:2006fy} but with their
parameters $V_L$, $V_R$, and $g_{L/R}$ being replaced by $f_1^{\rm SM}+\delta\! f_1^L$,
$f_1^R$, and $-f_2^{L/R}$ in our notations. The total decay width is derived
as the summation of the partial decay widths under the above assumption on the top-decay
channel.

The corresponding $W$-boson helicity fractions have been measured in Tevatron and LHC
experiments~\cite{Shabalina:2013mya}. In this analysis, we take the following data as
our input information~\cite{Khachatryan:2014vma}
\begin{equation}\label{eq:Frac_data}
\begin{split}
 &F_L^t=0.298\pm 0.028 (\rm stat.)\pm 0.032(\rm syst.),\\
 &F_0^t=0.720\pm 0.039 (\rm stat.)\pm 0.037 (\rm syst.),\\
 &F_R^t=-0.018\pm 0.019 (\rm stat.)\pm0.011(\rm syst.),
\end{split}
\end{equation}
and the total decay width of the top quark~\cite{Khachatryan:2014nda}
\begin{equation}\label{eq:total_w}
{\mit\Gamma}^t = 1.36\pm 0.02({\rm stat.})^{+0.14}_{-0.11}({\rm syst.}) ~~{\rm GeV},
\footnote{Since it is not easy to handle an asymmetric error like this in the error propagation,
    we use ${\mit\Gamma}^t = 1.36\pm 0.02({\rm stat.})\pm 0.14({\rm syst.}) ~{\rm GeV}$,
    the one symmetrized by adopting the larger (i.e., $+0.14$) in this systematic error,
    in the following calculation.}
\end{equation}
to get constraints on $\delta\! f_1^L$, $f_1^R$ and $f_{2}^{L/R}$.

We are, however, going to utilize the total and partial decay widths instead of using the above
$W$-boson helicity
fractions directly. This is because the fraction, defined by the ratio of the partial width 
to the total width, could reproduce experimental results in the case that the numerator
(i.e. partial width) and the denominator (i.e. total width) balance each other out, even if
they are both out of experimentally-allowed ranges. Therefore, we derive the partial decay widths
combining Eq.(\ref{eq:Frac_data}) and Eq.(\ref{eq:total_w}) as
\begin{equation}\label{eq:gamma_eff}
\begin{split}
 &{\mit\Gamma}_L^{t*}=0.405\pm 0.072~{\rm GeV},\\
 &{\mit\Gamma}_0^{t*}=0.979\pm 0.125 ~{\rm GeV},\\
 &{\mit\Gamma}_R^{t*}=-0.024\pm0.030~{\rm GeV},
\end{split}
\end{equation}
and use them as input data in our analyses.\footnote{The lower value of ${\mit\Gamma}_R^{t*}$ 
    is set to be zero in the actual calculation  because the decay width should not be
    a negative quantity.}

As mentioned, we handle the real and imaginary parts of all the non-standard couplings
independently and at the same time, that is, we are going to carry out a full eight-parameter
analysis. More specifically, we compare our input data (\ref{eq:total_w}) and (\ref{eq:gamma_eff})
with the corresponding formulas by varying each parameter in steps of 0.05, and explore the
allowed parameter space. We express the results by presenting the maximum and minimum values
of each parameter in the following.


At first, the result in the case that all the non-standard couplings are independent complex
numbers is shown in Table~\ref{tab:mgcase}. We there find that the constraints
on each couplings are not very strong.\footnote{Note that the results have an error of about
    0.05 because of our computational precision.}\
Thus even if each coupling is large, the experimental data can be reproduced
as a result of cancellations among the contributions from some of the couplings.
In particular, the constraint on $\delta\! f_1^L$ is weaker than on the other couplings.
It might seem strange that the contribution from the standard-model coupling $f^{\rm SM}_1$ 
is diminished by its extended coupling $\delta\! f_1^L$
but we of course cannot get rid of such a possibility.

\vspace{0.45cm}

\begin{table}[H]
\centering
\caption{The allowed maximum and minimum values of non-standard-top-decay couplings in the case
that all the couplings are dealt with as free parameters.}
\label{tab:mgcase}
\vspace*{0.4cm}
\begin{tabular}{c|cc|cc|cc|cc}
& \multicolumn{2}{c|}{$\delta\! f_1^L$}& \multicolumn{2}{c|}{$f_1^R$}
& \multicolumn{2}{c|}{$f_2^L$}& \multicolumn{2}{c}{$f_2^R$}
\\ \cline{2-9} 
& Re($\delta\! f_1^L$)
&\hspace*{-0.4cm} Im($\delta\! f_1^L$) & Re($f_1^R$)
&\hspace*{-0.4cm} Im($f_1^R$) & Re($f_2^L$)
&\hspace*{-0.4cm} Im($f_2^L$) & Re($f_2^R$)
&\hspace*{-0.4cm} Im($f_2^R$)
\\ \cline{1-9}
Min. & $-2.55$           &\hspace*{-0.4cm} $-1.55$           & $-1.30$           
&\hspace*{-0.4cm} $-1.30$           & $-0.65$           
&\hspace*{-0.4cm} $-0.65$           & $-1.20$            
&\hspace*{-0.4cm} $-1.20$           \\ 
Max. & $\phantom{-}0.55$ &\hspace*{-0.4cm} $\phantom{-}1.55$ & $\phantom{-}1.30$ 
&\hspace*{-0.4cm} $\phantom{-}1.30$ & $\phantom{-}0.65$ 
&\hspace*{-0.4cm} $\phantom{-}0.65$  & $\phantom{-}1.20$  
&\hspace*{-0.4cm} $\phantom{-}1.20$
\end{tabular}
\end{table}

\vspace{0.45cm}

Having these results, we then have considered the cases where Re$(\delta\! f_1^L)=0$ and also
Re$(\delta\! f_1^L)={\rm Im} (\delta\! f_1^L)=0$, and performed the same estimation for each
case. Their results are shown in Table~\ref{tab:7paraA} and Table~\ref{tab:tab:6paraA}.
As we see there, if the standard $V-A$ interaction, i.e., the $f_1^{\rm SM}$ term, is not
affected by $\delta\! f_1^L$, constraints on the remaining couplings get a bit stronger.
Moreover, it is remarkable that the allowed region of Re($f_2^R$) has become largely asymmetric
and the upper limits are both zero, which seems to indicate that a negative Re($f_2^R$) (in our
notation) is favored.

\vspace{0.45cm}

\begin{table}[H]
\centering
\caption{The allowed maximum and minimum values of non-standard-top-decay couplings in the case
that all the couplings are dealt with as free parameters
except for Re$(\delta\! f_1^L)$ being set to be zero.}
\label{tab:7paraA}
\vspace*{0.4cm}
\begin{tabular}{c|c|cc|cc|cc}
& {$\delta\! f_1^L$}& \multicolumn{2}{c|}{$f_1^R$}& \multicolumn{2}{c|}{$f_2^L$}
& \multicolumn{2}{c}{$f_2^R$}             \\ \cline{2-8}
&  Im($\delta\! f_1^L$) & Re($f_1^R$) & Im($f_1^R$) & Re($f_2^L$) &Im($f_2^L$) 
& Re($f_2^R$) & Im($f_2^R$) \\  \cline{1-8}
Min. &  $-1.20$  & $-1.10$ & $-1.10$ & $-0.50$ & $-0.55$ & $-0.95$ & $-1.00$ \\ 
Max. &  $\phantom{-}1.20$  & $\phantom{-}1.05$ & $\phantom{-}1.10$ & $\phantom{-}0.55$ 
     & $\phantom{-}0.55$   & $\phantom{-}0.00$ & $\phantom{-}1.00$
\end{tabular}
\end{table}
\begin{table}[H]
\centering
\caption{The allowed maximum and minimum values of non-standard-top-decay couplings in the case
that all the couplings are dealt with as free parameters except for Re$(\delta\! f_1^L)$ and
Im$(\delta\! f_1^L)$ both being set to be zero.}
\label{tab:tab:6paraA}
\vspace*{0.4cm}
\begin{tabular}{c|cc|cc|cc}
& \multicolumn{2}{c|}{$f_1^R$}& \multicolumn{2}{c|}{$f_2^L$}
& \multicolumn{2}{c}{$f_2^R$} \\ \cline{2-7} 
&  Re($f_1^R$) & Im($f_1^R$) & Re($f_2^L$) &Im($f_2^L$) & Re($f_2^R$) & Im($f_2^R$) \\ \cline{1-7}
Min. &  $-1.10 $ & $-1.10$ & $-0.50$ & $-0.55$ & $-0.95$ & $-0.45$  \\ 
Max. &  $\phantom{-}1.05$  & $\phantom{-}1.10$ & $\phantom{-}0.55$ & $\phantom{-}0.55$
     & $\phantom{-}0.00$   & $ \phantom{-}0.45$
\end{tabular}
\end{table}

\vspace{0.45cm}

Some comments should be mentioned on what we have obtained: The above asymmetric result is
not surprising because Re($f_2^R$) produces the only term which can interfere with the
standard-model coupling even when the $b$-quark is treated as a massless particle, that is,
we have a term proportional to this coupling in ${\mit\Gamma}^t_{L,0,R}$. Therefore,
the sign of Re($f_2^R$), if any, could be
determined from the measurable decay widths and/or $W$-boson helicity fractions in the near
future. On the other hand, let us not forget that an error around 0.05 is included in our
calculations, concerning the upper (and lower) bound. Finally, all the allowed parameter
spaces contain the standard-model solution, i.e., $\delta\! f_1^L=f_1^R=f_2^{L,R}=0$, which
means there is no new-physics signal yet in the quantities studied here.\footnote{Some
    non-vanishing contributions to these parameters are also made via
    standard-model radiative corrections, see \cite{Li:1990qf,GonzalezSprinberg:2011kx}.}

To summarize, we have studied possible non-standard $tbW$ interactions and found that
the present data are consistent with the standard-model predictions but there is some
non-negligible space left for possible non-standard couplings, too. We have derived
the maximum and minimum values of those couplings allowed by the present experimental
data of the total and partial decay widths by varying all the couplings independently
at the same time.

To be more specific, the conceivable non-standard-top-decay couplings are classified
into eight types if we treat all the coupling constants as complex numbers. In that case,
the allowed regions of those couplings are not that small yet because cancellations could
happen between the contributions originated from those couplings. On the other hand, if
we assume that $f_1^L$ does not include any non-standard contribution, the resultant
constraints on the other non-standard couplings, especially $f_2^R$, become a bit stronger,
although their allowed ranges are not such tiny that we can drop their quadratic terms
easily. These results tell us that we should be very careful when taking the linear
approximation on those non-standard $tbW$ couplings.

\vspace{0.4cm}

%
\secnon{Acknowledgments}
%
We would like to thank Akira Uejima for his quite helpful advice on numerical analyses
using the parallel-computing procedure.
This work was partly supported by the Grant-in-Aid for Scientific Research 
No. 22540284 from the Japan Society for the Promotion of Science.
Part of the algebraic and numerical calculations were carried out on the computer
system at Yukawa Institute for Theoretical Physics (YITP), Kyoto University.
\baselineskip=20pt plus 0.1pt minus 0.1pt

\vspace*{0.8cm}



\begin{thebibliography}{99}
%
\bibitem{Aad:2012tfa}
  G.~Aad {\it et al.} [ATLAS Collaboration],
  Phys.\ Lett.\ B {\bf 716} (2012) 1
  (arXiv:1207.7214 [hep-ex]).
%
\bibitem{Chatrchyan:2012xdj}
  S.~Chatrchyan {\it et al.} [CMS Collaboration],
  Phys.\ Lett.\ B {\bf 716} (2012) 30
  (arXiv:1207.7235 [hep-ex]).
%
\bibitem{Atwood:2000tu}
  D.~Atwood, S.~Bar-Shalom, G.~Eilam and A.~Soni,
  Phys.\ Rept.\  {\bf 347} (2001) 1
  [hep-ph/0006032].
%
\bibitem{Kamenik:2011wt}
  J.F.~Kamenik, J.~Shu and J.~Zupan,
  Eur.\ Phys.\ J.\ C {\bf 72} (2012) 2102
  (arXiv:1107.5257 [hep-ph]).
%
\bibitem{Tait:2015usa}
  T.M.P.~Tait,
  arXiv:1502.07029 [hep-ph].
%
\bibitem{LHC} LHC website: {\tt http://public.web.cern.ch/public/en/LHC/LHC-en.html}
%
\bibitem{Buchmuller:1985jz}
  W.~Buchmuller and D.~Wyler,
  Nucl.\ Phys.\ B {\bf 268} (1986) 621.
%
\bibitem{Arzt:1994gp}
  C.~Arzt, M.B.~Einhorn and J.~Wudka,
  Nucl.\ Phys.\ B {\bf 433} (1995) 41
  [hep-ph/9405214].
%
\bibitem{AguilarSaavedra:2008zc}
  J.A.~Aguilar-Saavedra,
  Nucl.\ Phys.\ B {\bf 812} (2009) 181
  (arXiv:0811.3842 [hep-ph]).
%
\bibitem{Grzadkowski:2010es}
  B.~Grzadkowski, M.~Iskrzynski, M.~Misiak and J.~Rosiek,
  JHEP {\bf 1010} (2010) 085
  (arXiv:1008.4884 [hep-ph]).
%
\bibitem{Bigi:1980az}
  I.I.Y.~Bigi and H.~Krasemann,
  Z.\ Phys.\ C {\bf 7} (1981) 127.
%
\bibitem{Bigi:1986jk}
  I.I.Y.~Bigi, Y.L.~Dokshitzer, V.A.~Khoze, J.H.~Kuhn and P.M.~Zerwas,
  Phys.\ Lett.\ B {\bf 181} (1986) 157.
%
\bibitem{Kane:1991bg}
  G.L.~Kane, G.A.~Ladinsky and C.P.~Yuan,
  Phys.\ Rev.\ D {\bf 45} (1992) 124.
%
\bibitem{Malkawi:1994tg}
  E.~Malkawi and C.P.~Yuan,
  Phys.\ Rev.\ D {\bf 50} (1994) 4462
  [hep-ph/9405322].
%
\bibitem{Carlson:1994bg}
  D.O.~Carlson, E.~Malkawi and C.P.~Yuan,
  Phys.\ Lett.\ B {\bf 337} (1994) 145
  [hep-ph/9405277].
%
\bibitem{Whisnant:1997qu}
  K.~Whisnant, J.M.~Yang, B.L.~Young and X.~Zhang,
  Phys.\ Rev.\ D {\bf 56} (1997) 467
  [hep-ph/9702305].
%
\bibitem{Yang:1997iv}
  J.M.~Yang and B.L.~Young,
  Phys.\ Rev.\ D {\bf 56} (1997) 5907
  [hep-ph/9703463].
%
\bibitem{Cao:1998at}
  J.J.~Cao, J.X.~Wang, J.M.~Yang, B.L.~Young and X.m.~Zhang,
  Phys.\ Rev.\ D {\bf 58} (1998) 094004
  [hep-ph/9804343].
%
\bibitem{Hikasa:1998wx}
  K.I.~Hikasa, K.~Whisnant, J.M.~Yang and B.L.~Young,
  Phys.\ Rev.\ D {\bf 58} (1998) 114003
  [hep-ph/9806401].
%
\bibitem{Larios:1999au}
  F.~Larios, M.A.~Perez and C.P.~Yuan,
  Phys.\ Lett.\ B {\bf 457} (1999) 334
  [hep-ph/9903394].
%
\bibitem{Espriu:2001vj}
  D.~Espriu and J.~Manzano,
  Phys.\ Rev.\ D {\bf 65} (2002) 073005
  [hep-ph/0107112].
%
\bibitem{AguilarSaavedra:2006fy}
  J.A.~Aguilar-Saavedra, J.~Carvalho, N.F.~Castro, F.~Veloso and A.~Onofre,
  Eur.\ Phys.\ J.\ C {\bf 50} (2007) 519
  [hep-ph/0605190].
%
\bibitem{Batra:2006iq}
  P.~Batra and T.M.P.~Tait,
  Phys.\ Rev.\ D {\bf 74} (2006) 054021
  [hep-ph/0606068].
%
\bibitem{Cao:2007ea}
  Q.H.~Cao, J.~Wudka and C.-P.~Yuan,
  Phys.\ Lett.\ B {\bf 658} (2007) 50
  (arXiv:0704.2809 [hep-ph]).
%
\bibitem{Berger:2009hi}
  E.L.~Berger, Q.H.~Cao and I.~Low,
  Phys.\ Rev.\ D {\bf 80} (2009) 074020
  (arXiv:0907.2191 [hep-ph]).
%
\bibitem{AguilarSaavedra:2010nx}
  J.A.~Aguilar-Saavedra and J.~Bernabeu,
  Nucl.\ Phys.\ B {\bf 840} (2010) 349
  (arXiv:1005.5382 [hep-ph]).
%
\bibitem{Zhang:2010dr}
  C.~Zhang and S.~Willenbrock,
  Phys.\ Rev.\ D {\bf 83} (2011) 034006
  (arXiv:1008.3869 [hep-ph]).
%
\bibitem{AguilarSaavedra:2011ct}
  J.A.~Aguilar-Saavedra, N.F.~Castro and A.~Onofre,
  Phys.\ Rev.\ D {\bf 83} (2011) 117301
  (arXiv:1105.0117 [hep-ph]).
%
\bibitem{Bach:2012fb}
  F.~Bach and T.~Ohl,
  Phys.\ Rev.\ D {\bf 86} (2012) 114026
  (arXiv:1209.4564 [hep-ph]).
%
\bibitem{Dutta:2013mva}
  S.~Dutta, A.~Goyal, M.~Kumar and B.~Mellado,
  arXiv:1307.1688 [hep-ph].
%
\bibitem{Prasath:2014mfa}
  A.V.~Prasath, R.M.~Godbole and S.D.~Rindani,
  Eur.\ Phys.\ J.\ C {\bf 75} (2015) 9, 402
  (arXiv:1405.1264 [hep-ph]).
%
\bibitem{Fabbrichesi:2014wva}
  M.~Fabbrichesi, M.~Pinamonti and A.~Tonero,
  Eur.\ Phys.\ J.\ C {\bf 74} (2014) 12,  3193
  (arXiv:1406.5393 [hep-ph]).
%
\bibitem{Bernardo:2014vha}
  C.~Bernardo, N.F.~Castro, M.C.~N.~Fiolhais, H.~Goncalves, A.G.C.~Guerra, M.~Oliveira and A.~Onofre,
  Phys.\ Rev.\ D {\bf 90} (2014) 11, 113007
  (arXiv:1408.7063 [hep-ph]).
%
\bibitem{Cao:2015doa}
  Q.H.~Cao, B.~Yan, J.H.~Yu and C.~Zhang,
  arXiv:1504.03785 [hep-ph].
%
\bibitem{Rindani:2015vya}
  S.D.~Rindani, P.~Sharma and A.W.~Thomas,
  JHEP10 (2015) 180
  (arXiv:1507.08385 [hep-ph]).
%
\bibitem{Aguilar:2015vsa}
  R.R.~Aguilar, A.O.~Bouzas and F.~Larios,
  arXiv:1509.06431 [hep-ph].
%
\bibitem{HIOKI:2011xx}
  Z.~Hioki and K.~Ohkuma,
  Phys.\ Rev.\ D {\bf 83} (2011) 114045
  (arXiv:1104.1221 [hep-ph]).
%
\bibitem{Hioki:2012vn}
  Z.~Hioki and K.~Ohkuma,
  Phys.\ Lett.\ B {\bf 716} (2012) 310
  (arXiv:1206.2413 [hep-ph]).
%
\bibitem{Hioki:2014eca}
  Z.~Hioki and K.~Ohkuma,
  Phys.\ Lett.\ B {\bf 736} (2014) 1
  (arXiv:1406.2475 [hep-ph]).
%
\bibitem{Shabalina:2013mya}
  E.~Shabalina [CDF and D0 and ATLAS and CMS Collaborations],
  EPJ Web Conf.\  {\bf 49} (2013) 07002.
%
\bibitem{Khachatryan:2014vma}
  V.~Khachatryan {\it et al.} [CMS Collaboration],
  JHEP {\bf 1501} (2015) 053
  (arXiv:1410.1154 [hep-ex]).
%
\bibitem{Khachatryan:2014nda}
  V.~Khachatryan {\it et al.} [CMS Collaboration],
  Phys.\ Lett.\ B {\bf 736} (2014) 33
  (arXiv:1404.2292 [hep-ex]).
%
\bibitem{Li:1990qf}
  C.S.~Li, R.J.~Oakes and T.C.~Yuan,
  Phys.\ Rev.\ D {\bf 43} (1991) 3759.
%
\bibitem{GonzalezSprinberg:2011kx}
  G.A.~Gonzalez-Sprinberg, R.~Martinez and J.~Vidal,
  JHEP {\bf 1107} (2011) 094 [JHEP {\bf 1305} (2013) 117]
  (arXiv:1105.5601 [hep-ph]).
\end{thebibliography}
\end{document}